\RequirePackage{lineno}
\documentclass[%
amsmath,amssymb,
aps,
prl,
twocolumn,
superscriptaddress,
 longbibliography,
]{revtex4}

\usepackage{verbatim}
\usepackage{graphicx}
\usepackage{dcolumn}
\usepackage{bm}
\usepackage{hyperref}
\usepackage{slashed}
\usepackage{lineno}

\def\simge{\mathrel{%
   \rlap{\raise 0.511ex \hbox{$>$}}{\lower 0.511ex \hbox{$\sim$}}}}
\def\simle{\mathrel{
   \rlap{\raise 0.511ex \hbox{$<$}}{\lower 0.511ex \hbox{$\sim$}}}}

\newcommand{\MET}{\mbox{$E\kern-0.50em\raise0.10ex\hbox{/}_{T}$}}
\newcommand{\METsig}{\mbox{$E\kern-0.50em\raise0.10ex\hbox{/}_{T,\mathrm{sig}}$}}
\newcommand{\METrel}{\mbox{$\min E\kern-0.50em\raise0.10ex\hbox{/}_{T,\mathrm{rel}}$}}

\begin{document}

\title{\boldmath An inclusive search for the Higgs boson in the four-lepton final state at CDF} 

\affiliation{Institute of Physics, Academia Sinica, Taipei, Taiwan 11529, Republic of China}
\affiliation{Argonne National Laboratory, Argonne, Illinois 60439, USA}
\affiliation{University of Athens, 157 71 Athens, Greece}
\affiliation{Institut de Fisica d'Altes Energies, ICREA, Universitat Autonoma de Barcelona, E-08193, Bellaterra (Barcelona), Spain}
\affiliation{Baylor University, Waco, Texas 76798, USA}
\affiliation{Istituto Nazionale di Fisica Nucleare Bologna, $^{ee}$University of Bologna, I-40127 Bologna, Italy}
\affiliation{University of California, Davis, Davis, California 95616, USA}
\affiliation{University of California, Los Angeles, Los Angeles, California 90024, USA}
\affiliation{Instituto de Fisica de Cantabria, CSIC-University of Cantabria, 39005 Santander, Spain}
\affiliation{Carnegie Mellon University, Pittsburgh, Pennsylvania 15213, USA}
\affiliation{Enrico Fermi Institute, University of Chicago, Chicago, Illinois 60637, USA}
\affiliation{Comenius University, 842 48 Bratislava, Slovakia; Institute of Experimental Physics, 040 01 Kosice, Slovakia}
\affiliation{Joint Institute for Nuclear Research, RU-141980 Dubna, Russia}
\affiliation{Duke University, Durham, North Carolina 27708, USA}
\affiliation{Fermi National Accelerator Laboratory, Batavia, Illinois 60510, USA}
\affiliation{University of Florida, Gainesville, Florida 32611, USA}
\affiliation{Laboratori Nazionali di Frascati, Istituto Nazionale di Fisica Nucleare, I-00044 Frascati, Italy}
\affiliation{University of Geneva, CH-1211 Geneva 4, Switzerland}
\affiliation{Glasgow University, Glasgow G12 8QQ, United Kingdom}
\affiliation{Harvard University, Cambridge, Massachusetts 02138, USA}
\affiliation{Division of High Energy Physics, Department of Physics, University of Helsinki and Helsinki Institute of Physics, FIN-00014, Helsinki, Finland}
\affiliation{University of Illinois, Urbana, Illinois 61801, USA}
\affiliation{The Johns Hopkins University, Baltimore, Maryland 21218, USA}
\affiliation{Institut f\"{u}r Experimentelle Kernphysik, Karlsruhe Institute of Technology, D-76131 Karlsruhe, Germany}
\affiliation{Center for High Energy Physics: Kyungpook National University, Daegu 702-701, Korea; Seoul National University, Seoul 151-742, Korea; Sungkyunkwan University, Suwon 440-746, Korea; Korea Institute of Science and Technology Information, Daejeon 305-806, Korea; Chonnam National University, Gwangju 500-757, Korea; Chonbuk National University, Jeonju 561-756, Korea}
\affiliation{Ernest Orlando Lawrence Berkeley National Laboratory, Berkeley, California 94720, USA}
\affiliation{University of Liverpool, Liverpool L69 7ZE, United Kingdom}
\affiliation{University College London, London WC1E 6BT, United Kingdom}
\affiliation{Centro de Investigaciones Energeticas Medioambientales y Tecnologicas, E-28040 Madrid, Spain}
\affiliation{Massachusetts Institute of Technology, Cambridge, Massachusetts 02139, USA}
\affiliation{Institute of Particle Physics: McGill University, Montr\'{e}al, Qu\'{e}bec, Canada H3A~2T8; Simon Fraser University, Burnaby, British Columbia, Canada V5A~1S6; University of Toronto, Toronto, Ontario, Canada M5S~1A7; and TRIUMF, Vancouver, British Columbia, Canada V6T~2A3}
\affiliation{University of Michigan, Ann Arbor, Michigan 48109, USA}
\affiliation{Michigan State University, East Lansing, Michigan 48824, USA}
\affiliation{Institution for Theoretical and Experimental Physics, ITEP, Moscow 117259, Russia}
\affiliation{University of New Mexico, Albuquerque, New Mexico 87131, USA}
\affiliation{The Ohio State University, Columbus, Ohio 43210, USA}
\affiliation{Okayama University, Okayama 700-8530, Japan}
\affiliation{Osaka City University, Osaka 588, Japan}
\affiliation{University of Oxford, Oxford OX1 3RH, United Kingdom}
\affiliation{Istituto Nazionale di Fisica Nucleare, Sezione di Padova-Trento, $^{ff}$University of Padova, I-35131 Padova, Italy}
\affiliation{University of Pennsylvania, Philadelphia, Pennsylvania 19104, USA}
\affiliation{Istituto Nazionale di Fisica Nucleare Pisa, $^{gg}$University of Pisa, $^{hh}$University of Siena and $^{ii}$Scuola Normale Superiore, I-56127 Pisa, Italy}
\affiliation{University of Pittsburgh, Pittsburgh, Pennsylvania 15260, USA}
\affiliation{Purdue University, West Lafayette, Indiana 47907, USA}
\affiliation{University of Rochester, Rochester, New York 14627, USA}
\affiliation{The Rockefeller University, New York, New York 10065, USA}
\affiliation{Istituto Nazionale di Fisica Nucleare, Sezione di Roma 1, $^{jj}$Sapienza Universit\`{a} di Roma, I-00185 Roma, Italy}
\affiliation{Rutgers University, Piscataway, New Jersey 08855, USA}
\affiliation{Texas A\&M University, College Station, Texas 77843, USA}
\affiliation{Istituto Nazionale di Fisica Nucleare Trieste/Udine, I-34100 Trieste, $^{kk}$University of Udine, I-33100 Udine, Italy}
\affiliation{University of Tsukuba, Tsukuba, Ibaraki 305, Japan}
\affiliation{Tufts University, Medford, Massachusetts 02155, USA}
\affiliation{University of Virginia, Charlottesville, Virginia 22906, USA}
\affiliation{Waseda University, Tokyo 169, Japan}
\affiliation{Wayne State University, Detroit, Michigan 48201, USA}
\affiliation{University of Wisconsin, Madison, Wisconsin 53706, USA}
\affiliation{Yale University, New Haven, Connecticut 06520, USA}

\author{T.~Aaltonen}
\affiliation{Division of High Energy Physics, Department of Physics, University of Helsinki and Helsinki Institute of Physics, FIN-00014, Helsinki, Finland}
\author{B.~\'{A}lvarez~Gonz\'{a}lez$^z$}
\affiliation{Instituto de Fisica de Cantabria, CSIC-University of Cantabria, 39005 Santander, Spain}
\author{S.~Amerio}
\affiliation{Istituto Nazionale di Fisica Nucleare, Sezione di Padova-Trento, $^{ff}$University of Padova, I-35131 Padova, Italy}
\author{D.~Amidei}
\affiliation{University of Michigan, Ann Arbor, Michigan 48109, USA}
\author{A.~Anastassov$^x$}
\affiliation{Fermi National Accelerator Laboratory, Batavia, Illinois 60510, USA}
\author{A.~Annovi}
\affiliation{Laboratori Nazionali di Frascati, Istituto Nazionale di Fisica Nucleare, I-00044 Frascati, Italy}
\author{J.~Antos}
\affiliation{Comenius University, 842 48 Bratislava, Slovakia; Institute of Experimental Physics, 040 01 Kosice, Slovakia}
\author{G.~Apollinari}
\affiliation{Fermi National Accelerator Laboratory, Batavia, Illinois 60510, USA}
\author{J.A.~Appel}
\affiliation{Fermi National Accelerator Laboratory, Batavia, Illinois 60510, USA}
\author{T.~Arisawa}
\affiliation{Waseda University, Tokyo 169, Japan}
% ...

\collaboration{CDF Collaboration\footnote{With visitors from
$^a$Istituto Nazionale di Fisica Nucleare, Sezione di Cagliari, 09042 Monserrato (Cagliari), Italy,
$^b$University of CA Irvine, Irvine, CA 92697, USA,
$^c$University of CA Santa Barbara, Santa Barbara, CA 93106, USA,
$^d$University of CA Santa Cruz, Santa Cruz, CA 95064, USA,
$^e$Institute of Physics, Academy of Sciences of the Czech Republic, Czech Republic,
$^f$CERN, CH-1211 Geneva, Switzerland,
$^g$Cornell University, Ithaca, NY 14853, USA,
$^h$University of Cyprus, Nicosia CY-1678, Cyprus,
$^i$Office of Science, U.S. Department of Energy, Washington, DC 20585, USA,
$^j$University College Dublin, Dublin 4, Ireland,
$^k$ETH, 8092 Zurich, Switzerland,
$^l$University of Fukui, Fukui City, Fukui Prefecture, Japan 910-0017,
$^m$Universidad Iberoamericana, Mexico D.F., Mexico,
$^n$University of Iowa, Iowa City, IA 52242, USA,
$^o$Kinki University, Higashi-Osaka City, Japan 577-8502,
$^p$Kansas State University, Manhattan, KS 66506, USA,
$^q$Ewha Womans University, Seoul, 120-750, Korea,
$^r$University of Manchester, Manchester M13 9PL, United Kingdom,
$^s$Queen Mary, University of London, London, E1 4NS, United Kingdom,
$^t$University of Melbourne, Victoria 3010, Australia,
$^u$Muons, Inc., Batavia, IL 60510, USA,
$^v$Nagasaki Institute of Applied Science, Nagasaki, Japan,
$^w$National Research Nuclear University, Moscow, Russia,
$^x$Northwestern University, Evanston, IL 60208, USA,
$^y$University of Notre Dame, Notre Dame, IN 46556, USA,
$^z$Universidad de Oviedo, E-33007 Oviedo, Spain,
$^{aa}$CNRS-IN2P3, Paris, F-75205 France,
$^{bb}$Texas Tech University, Lubbock, TX 79609, USA,
$^{cc}$Universidad Tecnica Federico Santa Maria, 110v Valparaiso, Chile,
$^{dd}$Yarmouk University, Irbid 211-63, Jordan,
}}
\noaffiliation


\begin{abstract}
An inclusive search for the standard model Higgs boson using the four-lepton final state in 
proton-antiproton collisions produced by the Tevatron at  $\sqrt{s}=1.96$ TeV is conducted. The data are
recorded by the CDF II detector and correspond to an integrated luminosity of 9.7 fb$^{-1}$.
Three distinct Higgs decay modes, namely {\it ZZ}, {\it WW}, and $\tau\tau$, are simultaneously probed. 
Nine potential signal events are selected and found to be consistent with the background expectation.
We set a 95\% credibility limit on the production cross section times the branching ratio and subsequent
decay to the four lepton final state  for hypothetical Higgs boson masses between $120\,\mathrm{GeV}/c^2$ and $300\,\mathrm{GeV}/c^2$.
\end{abstract}

\pacs{PACS}

\maketitle

\date{\today}

The vector gauge bosons mediating the weak force, the $W$ and $Z$, are 
massive. Within the standard model of particle physics (SM) their masses 
arise through spontaneous symmetry 
breaking (SSB) of the electroweak symmetry~\cite{Glashow:1961tr,PhysRevLett.19.1264,sm_salam} through 
the introduction of a scalar field that maintains the gauge invariance of the theory~\cite{Higgs:1964ia,Higgs:1964pj,Guralnik:1964eu,Englert:1964et}. 
Fundamental fermions may acquire mass through Yukawa couplings with this field. 
Quantization of the field necessitates the existence of an associated spin-0 particle, known as the Higgs boson,
the discovery of which would confirm the SM mechanism for electroweak SSB. 

Direct searches at LEP~\cite{Barate:2003sz}, combined with recent search results from the   
Tevatron~\cite{TEVNPH:2012ab} and LHC experiments~\cite{ATLAS:2012ae,Chatrchyan:2012tx}, exclude 
all potential SM Higgs masses outside the ranges  $116.6 - 119.4\,\mathrm{GeV}/c^2$ and $122.1 - 127\,\mathrm{GeV}/c^2$. Although signal processes leading to four-lepton final states are not the most dominant within these mass ranges 
the inclusion of additional channels improves overall search sensitivity. 
Moreover, using the new techniques presented here, the search is simultaneously sensitive to multiple production and decay modes.
Also based on this inclusive approach, information regarding Higgs boson couplings to both fermions and bosons can be 
extracted, which could be useful to probe the SM or beyond SM nature of a hypothetical signal.

Searches for the SM Higgs boson to four leptons through the decay $ZZ^{(*)}$ have been published by ATLAS~\cite{Atl:2012hzz} and 
CMS~\cite{Chatrchyan:2012dg} using proton-proton collision data at a center of mass energy of 
$7\,\mathrm{TeV}$. 
In this Letter a search for the Higgs boson produced inclusively and decaying into final 
states containing either four electrons ($4e$), four muons ($4\mu$) or two electrons and 
two muons  ($2e2\mu$) is reported. Data from proton-antiproton collisions at center of mass energy 
$1.96~\mathrm{TeV}$, collected with the CDF~II detector \cite{Abulencia:2005ix} at the Tevatron and corresponding to $9.7\,\mathrm{fb^{-1}}$ of integrated
luminosity, are used.
Higgs boson decays to $Z$ boson pairs ($H\to$ {\it ZZ}) are the dominant contribution for most of the considered 
Higgs boson mass range. Owing to the full reconstruction of the final state, the  {\it ZZ} channel provides good search sensitivity in spite of the small potential signal yields because the resonance structure of the signal can be exploited to distinguish it from non-resonant backgrounds. Along with the direct Higgs production 
mechanisms of gluon-fusion ({\it ggH}) and fusion of vector bosons emitted from the incoming partons (VBF), the search  
is sensitive to associated Higgs boson production processes ({\it VH}). In the case of $ZH$ production, additional potential four-lepton event contributions originate from Higgs boson decays to $W$ boson pairs  ($ZH\to ZWW \to lll \nu l \nu$) and $\tau$-lepton pairs ($ZH\to Z\tau\tau \to lll\nu\nu l \nu\nu$). 
The detection of four leptons provides one of the cleanest signatures available at a hadron 
collider. 
Because of the small probability associated with mimicking the signature of isolated
lepton candidates in the detector, background contributions from ubiquitous multi-jet
production processes are negligible in this final state.

Separation of potential signal and background event contributions is obtained from the
observed four-lepton invariant mass ($m_{4\ell}$) and missing transverse energy spectra~\footnote{The \MET is the magnitude of the missing
transverse energy vector $\vec{\MET}$, which is defined as the opposite of the vector sum of the $E_T$ of all calorimetric towers,
corrected to produce the correct average calorimeter response to jets and to muons.}.
Sensitivity is significantly enhanced through the inclusion of \MET\ information since $H\to WW$ and $H\to\tau\tau$ decays 
to leptons result in final state neutrinos, while none are expected in background events. The inclusion of \MET\, significantly 
improves the search sensitivity at Higgs masses below 150 GeV$/c^2$, and represents the main advancement in analysis technique over previously published searches in this final state.


Our search is performed using the CDF II detector consisting 
of a solenoidal spectrometer with a 
silicon tracker and an open-cell drift chamber (COT) surrounded by 
calorimeters and muon detectors~\cite{Abulencia:2005ix}. The geometry is 
characterized using the azimuthal angle $\phi$ and the pseudorapidity $\eta$.
Transverse energy, $E_T$, is defined as
$E \sin \theta$, where $E$ is the energy of an electromagnetic 
or hadronic calorimeter energy cluster. Transverse momentum, $p_T$, 
is the track momentum component transverse to the beam line.

Electron candidates are identified by matching a central or forward track
to energy deposited within the calorimeter.
Muon candidates are formed from charged particle tracks matched to minimum ionizing energy deposition 
in the calorimeter, which may or may not be matched 
to track segments (\emph{stubs}) in the muon chambers situated behind the calorimeters.
Lepton reconstruction algorithms are well validated and described in detail elsewhere \cite{Abulencia:2005ix}.
Taus are included in this search only if they decay to electrons or muons.

Candidate leptons are separated into various
categories: electrons, muons, and isolated tracks that project to detector regions with 
insufficient calorimeter coverage for energy measurements. 
Electron candidates are distinguished by whether they are found in the central or forward calorimeters ($|\eta|>1.1$)
where only silicon tracking information is available. The electron selection relies on track quality, track-calorimeter 
matching, calorimeter energy, calorimeter profile shape, and isolation information.
Most muon candidates rely on direct detection in the muon chambers, which are
distinguished by their acceptance in pseudorapidity: central muon detectors ($|\eta|<0.6$), 
central muon extension detectors ($0.6<|\eta|<1.0$),  and  
the intermediate muon detector ($1.0<|\eta|<1.5$). 
Remaining muon candidates rely on track matches to energy deposits consistent with a minimum
ionizing charged particle in the central 
and forward electromagnetic calorimeters respectively, failing to have an 
associated stub in the muon sub-detectors.
Isolated tracks pointing to uninstrumented regions of the detector and satisfying high-quality
requirements are reconstructed as leptons of unidentified flavor.
All leptons are required to be isolated by imposing the condition that 
the sum of the transverse energy of the calorimeter towers in a cone of 
$\Delta R \equiv \sqrt{(\Delta\phi)^2+(\Delta\eta)^2}=0.4$
around the lepton is less than $10$\% of the electron $E_T$ (muon $p_T$). 
A similar isolation requirement is applied based on the reconstructed tracks in a $\Delta R=0.4$ cone
around the lepton candidate track.

The probability that a jet will be misidentified as a lepton is measured using jet-enriched samples
and is corrected for the contributions of leptons from $W$ and $Z$ boson decays.
An average value is obtained from samples collected with different single jet $E_T$
threshold requirements (20, 50, 70, and 100 GeV) and an uncertainty is assigned
based on the spread within the individual measurements.
The range of measured misidentification probabilities for the lepton categories, 
which vary according to $E_T$ or $p_T$, is 
$0.5\%-3\%$ (central electrons), 
$2\%-6\%$ (forward electrons), 
$0.5\%-4\%$ (central muons),
$0.5\%-2\%$ (extension muons),
$0.5\%-2\%$ (intermediate muons),
$0.5\%-6\%$ (calorimeter only muons),
and $0.5\%-3\%$ (isolated tracks).

Events for the analysis are collected using online event selection (trigger) requirements corresponding
to the presence of a single high-$E_T$ electron or high-$p_T$ muon.
The electron trigger requires an electromagnetic energy cluster in the central 
calorimeter with $E_T > 18$ GeV geometrically matching the direction of 
a charged particle reconstructed in the COT with $p_T > 8$ GeV/c. 
Muon triggers are based on track segments 
in the muon chambers geometrically matching the direction of a charged 
particle reconstructed in the COT with $p_T > 18$ GeV/c. 
Trigger efficiencies are measured using samples of leptonic Z decays~\cite{Abulencia:2005ix}. 
To ensure a uniform trigger efficiency over the lepton momentum spectra, 
an offline selection is applied to the lepton matched to the trigger object,
requiring $E_T > 20$ GeV ($p_T > 20$ GeV/c) for electrons (muons).

Additional charged electrons (muons) are required to have $E_T > 10$ GeV 
($p_T > 10$ GeV/c). Exactly four leptons are required, each separated from the others by 
$\Delta R \geq 0.1$. This analysis evolved 
from a CDF measurement of the {\it ZZ} production cross section in the four lepton final state~\cite{ZZXS}, where 
constraints on the invariant mass of opposite-sign same-flavor dilepton pairs were imposed in order
to explicitly reconstruct $Z$ bosons.
For Higgs boson masses smaller than $180\,\mathrm{GeV}/c^2$, at least one of the $Z$ bosons 
is off-shell and such requirements on the mass become inefficient. 
We therefore require dilepton pair masses to be between 20 and 140 GeV/$c^2$. 
In the final state where all leptons are of the same flavor, opposite-sign pairings are assigned based on the 
the smallest deviation of the reconstructed masses from the known $Z$ boson mass.
Because the backgrounds are modest, loose constraints on the mass improve the sensitivity of the search. 
Higgs boson production can also involve hadronic jets originating from the associated vector boson in $WH$ or $ZH$ production,
the additional quarks in the forward detector from VBF production, or initial state gluon radiation.
Therefore, we place no restriction on the number of reconstructed jets contained within each event.

The selected events consist primarily of the background 
from non-resonant diboson production of $Z^{*}/Z$-boson pairs ({\it ZZ}). 
Smaller contributions originate from non-fully leptonic {\it ZZ} decays and $Z\gamma$ production,
which lead to events with two or three real reconstructed leptons and one or two falsely
reconstructed leptons associated with the photon or additional hadronic jet activity.
The background from top-quark pair production is found to be negligible ($<$0.01 events).  

The acceptances, efficiencies and kinematic properties 
of the signal and background processes are determined from simulation. 
$\textsc{pythia}$ \cite{Sjostrand:2006za} is used to model all Higgs boson production 
processes and the non-resonant {\it ZZ} background.
The modeling of $Z\gamma$ is based on the simulation framework developed by Bauer and Berger~\cite{Baur:1992cd}. 
For all samples CTEQ5L parton distribution functions (PDFs) are used to model the momentum distribution of 
the initial-state partons \cite{Lai:1999wy}. 
The cross sections for each process are 
normalized to next-to-next-to-leading order (NNLO) calculations with logarithmic resummation 
for ggH~\cite{deFlorian:2009hc,Anastasiou:2008tj}, 
NNLO for VH~\cite{Assamagan:2004mu,Brein:2003wg,Ciccolini:2003jy}, and next-to-leading 
order (NLO) calculations for VBF~\cite{Assamagan:2004mu,Berger:2004pca}, {\it ZZ}~\cite{SMXS}, 
and $Z\gamma$~\cite{Baur:1998}. 

The response of the CDF II detector is modeled with a $\textsc{geant}$-based 
simulation~\cite{GEANT}. Efficiency corrections for the 
simulated CDF II detector response for leptons and photon 
conversions were determined using independent data samples.

The overall normalization of the estimated background from false leptons is derived using a 
data-driven approach while the shape of the distribution is derived from MC simulation. 
The total contribution from false lepton events is estimated using events with two or 
three leptons and additional jets that may be falsely reconstructed as leptons 
weighted by the measured jet-to-lepton misidentification probabilities.
Since the number of available three-lepton events in data is small, we
model kinematic distributions using a weighted sum of those obtained from the simulated
{\it ZZ} and $Z\gamma$ MC samples.

The kinematic distributions of {\it ZZ} events with falsely reconstructed leptons is assumed to be
the same as that for correctly reconstructed {\it ZZ} events.
We apply the same procedure to the simulated $Z\gamma$ sample to obtain the invariant mass 
distribution for events from this process, which are found to be well modeled by a Landau function.          
We model \MET\ in $Z\gamma$ events using the distribution from simulated {\it ZZ} events with falsely
reconstructed leptons, which was found to agree.

In the ranges
$50 < m_{4l}< 600\,\mathrm{GeV}/c^2$  and $0 < \MET<200\,\mathrm{GeV}$,
we estimate background contributions of $10.59 \pm 1.34$ {\it ZZ} events and $0.39 \pm 0.19$
events with falsely reconstructed leptons.  
For a SM Higgs boson with a mass of $125$ GeV/c$^2$, contributions of
$0.053$ ($ggH$), 
$0.003$ (VBF), 
$0.006$ ($WH$), and 
$0.089$ ($ZH$) events, 
yielding a total of $0.15 \pm 0.01$ events, are expected. 
The indicated uncertainties include statistical and systematic contributions added in quadrature.
Systematic uncertainties are described below.

\begin{figure}
   \includegraphics[width=0.88\columnwidth]{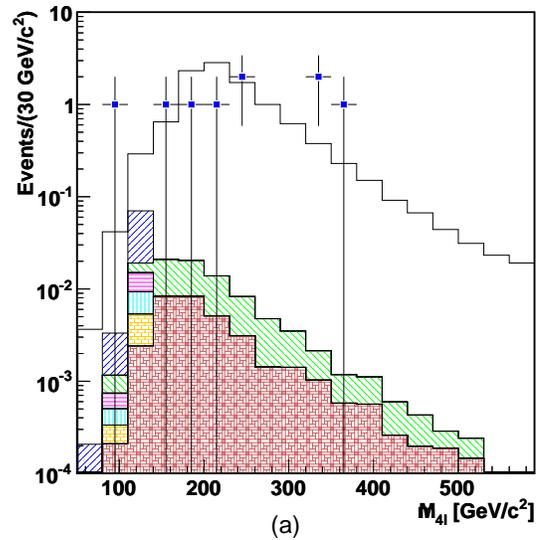} \\
   \includegraphics[width=0.88\columnwidth]{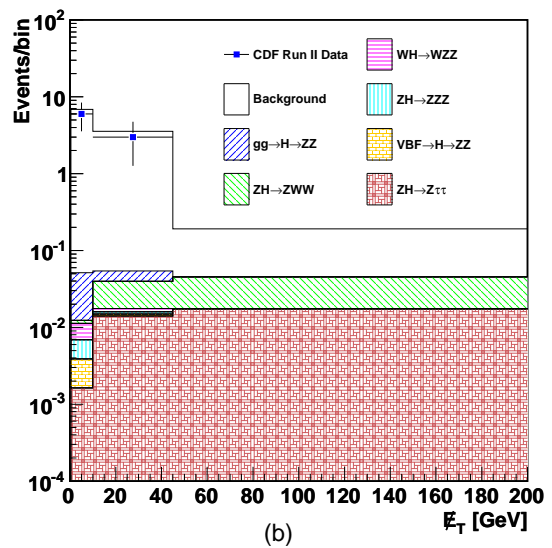} \\
  \caption{\label{fig:data} Distributions of the four-lepton invariant mass (a) and the missing transverse energy (b) in data. 
The combined estimated contribution from non-resonant {\it ZZ} production and fake leptons, denoted as background, is overlaid.
The potential contributions of the different Higgs production processes for $m_H=~$125 GeV/c$^2$ are stacked and also overlaid.}
\end{figure}

We observe a total of nine events, which is consistent with the rate expected from background sources only. 
The four-lepton invariant mass and \MET\, distributions are shown in Fig.~\ref{fig:data}(a) and 
Fig.~\ref{fig:data}(b) respectively, both overlaid with expected contributions from backgrounds 
and Higgs boson production for a mass of $125\,\mathrm{GeV}/c^2$, shown separately for 
each production process.
 
A variety of possible systematic effects were considered including those that affect 
the normalization and the shape of 
the kinematic distributions. The dominant systematic 
uncertainties are those on the theory predictions for the 
cross sections of signal and background processes. 
Systematic uncertainties 
associated with the MC affect Higgs signal, {\it ZZ}, and $Z\gamma$ acceptances 
determined from the  simulated event samples.

Uncertainties originating from lepton selection and 
trigger efficiency measurements are propagated through the acceptance 
calculation, leading to uncertainties of 3.6\% and 0.5\%, on 
the predicted signal and background event yields, respectively. In addition, all signal and background 
estimates obtained from simulation have an additional 5.9\% 
uncertainty originating from the measurement of the luminosity~\cite{Acosta:2002hx}. 

The $gg\to H$ cross-section has been computed at NNLO and  
next-to-next-to-leading log (NNLL) precision 
varying the renormalization and factorization scales and implementing different
correlated combinations of the MSTW2008 error PDFs together with allowed values of the
strong coupling constant ($\alpha_s$)~\cite{Catani:2007vq,Grazzini:2008tf},
yielding systematic uncertainties of $7.0\%$ and $7.7\%$, respectively.
Uncertainties on VBF and associated Higgs boson production, 
which account for about a quarter of the total Higgs boson events in our sample, are 
of $5\%$ and $10\%$, respectively~\cite{TeV4LHC}. 
A 3\% uncertainty was assigned on the branching fraction for $H \to ZZ$ and $H \to WW$, 
which are 100\% correlated, as well as a 3\% uncorrelated uncertainty on $H\to\tau\tau$~\cite{TeV4LHC}. 
The {\sc pythia} MC simulation for {\it ZZ} production that is used for determining acceptances 
is based on an expansion at LO; {\sc mcfm} \cite{mcfm} was used to estimate  
the potential difference in the acceptance arising from a full NLO simulation, 
estimated to be $\pm 2.5\%$ and assigned as a systematic uncertainty.
We assign a 10\% uncertainty on the {\it ZZ} cross-section based on the difference 
between LO and NLO~\cite{Campbell:1999ah} predictions.
The $Z\gamma$ component was parametrized using a Landau distribution, where an 
uncorrelated 50\% uncertainty was assigned to the yield prediction in each bin to 
fully cover the possible effects of mis-modeling. 
Misindentification probabilities were measured in several jet samples and the 
maximum spread between these measurements was assigned as a systematic uncertainty on the 
background estimation. Propagated through to the acceptance, 
this technique results in a 50\% variation in the estimate of the background from fake leptons. The missing transverse energy is 
scaled up and down by 20\% to account
for potential MC mis-modeling, and the modified shapes are taken as shape uncertainties on the
MC \MET\ distributions used as inputs to the limit calculation.

Only three \MET\ bins (0--15, 15--45, and 45--200 GeV) are used in the limit setting procedure.
The varied spacing was 
chosen as a compromise between search sensitivity and ensuring sufficient number of simulated events in each bin.

As a cross-check the distribution of the number of jets per event in data was compared to the
expectation from the {\it ZZ} MC simulation and found to be consistent.

\begin{figure}
  \begin{tabular}{c}
    \includegraphics[width=1.00\columnwidth]{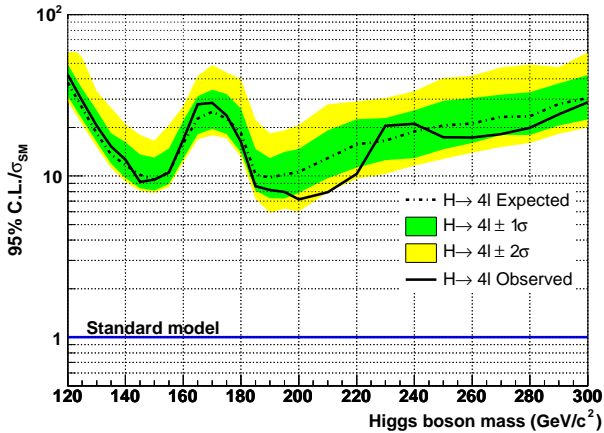} \\
  \end{tabular}
\caption{\label{fig:limits}
Expected and observed upper limits on the Higgs boson production relative to the SM prediction 
as a function of Higgs boson mass.
}
\end{figure}

Upper limits at the $95$\% credibility level (C.L.) are set
on the Higgs boson production cross section, $\sigma_H$, as a function of $m_H$. 
A Bayesian technique~\cite{Beringer:2012} was employed, where the posterior probability density was constructed 
from the joint Poisson probability of observing 
the data in each bin of the $m_{4l}-$\MET\ space, 
integrating over the uncertainties 
of the normalization parameters using Gaussian priors. 
A non-negative constant prior in the signal rate was assumed. The expected limit 
and associated one and two sigma bands are shown along 
with the observed limit in Table~\ref{tab:limits} and Fig.~\ref{fig:limits}. 
For a Higgs boson with $m_H=~$125 GeV/c$^2$ we expect a sensitivity of $26.5\times\sigma_{SM}$ at 95\% C.L. 
while we observe a limit of $29.3\times\sigma_{SM}$ at 95\% C.L. .
The results obtained in the investigated mass range are consistent with 
the observed event excess near 325 GeV/c$^2$ in the CDF search for high-mass resonances decaying to {\it ZZ}~\cite{Aaltonen:2011zq}. The analysis reported here is performed using standard CDF tracking algorithms while \cite{Aaltonen:2011zq} uses an alternative reconstruction.

\begin{table*}
  \caption{\label{tab:limits}Expected and observed upper limits on the Higgs boson production 
relative to the SM prediction for Higgs particle masses from 120 GeV/c$^2$ to 300 GeV/c$^2$.
}
\begin{ruledtabular}
\begin{center}
\tiny
\begin{tabular}{lrrrrrrrrrrrrrrrrrrr}
H$\to$4$\ell$   &         120 &         130 &         140 &         150 &         160 &         170 &         180 &         190 &         200 &         210 &         220 &         230 &         240 &         250 &         260 &         270 &         280 &         290 &         300 \\
\hline
   $\mathrm{Expected}/\sigma_{SM}$ &    38.0 &    18.3 &    11.7 &     9.4 &    16.0 &    25.1 &    18.5 &     9.8 &    10.6 &    12.9 &    15.7 &    16.6 &    18.9 &    20.5 &    21.1 &    23.2 &    23.5 &    28.0 &    30.5 \\
\hline
 $ \mathrm{Observed}/\sigma_{SM}$ &    42.4 &    20.5 &    12.6 &     9.5 &    16.8 &    28.5 &    16.3 &     8.2 &     7.2 &     7.9 &    10.3 &    20.5 &    21.1 &    17.4 &    17.3 &    18.2 &    19.9 &    24.1 &    28.6 \\
\end{tabular}

\end{center}
  \end{ruledtabular}    
\end{table*}

In $9.7~\mathrm{fb}^{-1}$ of integrated luminosity collected at the Tevatron,
no evidence for a Higgs boson signal is found in the mass range explored. 
Upper limits on the Higgs boson production cross section in the inclusive four-lepton 
final state relative to the SM expectation are obtained. 
Maximal search sensitivity is obtained not only in the high
mass region where both $Z$ bosons are produced on-shell but also in the lower 
mass region around $150\,\mathrm{GeV}/c^2$ where additional signal contributions 
from $ZH\to ZWW$ and $ZH\to Z\tau\tau$ improve the sensitivity by 15\%.


We thank the Fermilab staff and the technical staffs of the participating institutions for their vital contributions. This work was supported by the U.S. Department of Energy and National Science Foundation; the Italian Istituto Nazionale di Fisica Nucleare; the Ministry of Education, Culture, Sports, Science and Technology of Japan; the Natural Sciences and Engineering Research Council of Canada; the National Science Council of the Republic of China; the Swiss National Science Foundation; the A.P. Sloan Foundation; the Bundesministerium f\"ur Bildung und Forschung, Germany; the Korean World Class University Program, the National Research Foundation of Korea; the Science and Technology Facilities Council and the Royal Society, UK; the Russian Foundation for Basic Research; the Ministerio de Ciencia e Innovaci\'{o}n, and Programa Consolider-Ingenio 2010, Spain; the Slovak R\&D Agency; the Academy of Finland; and the Australian Research Council (ARC).

\bibliography{H4l_prl}

\begin{thebibliography}{35}
\expandafter\ifx\csname natexlab\endcsname\relax\def\natexlab#1{#1}\fi
\expandafter\ifx\csname bibnamefont\endcsname\relax
  \def\bibnamefont#1{#1}\fi
\expandafter\ifx\csname bibfnamefont\endcsname\relax
  \def\bibfnamefont#1{#1}\fi
\expandafter\ifx\csname citenamefont\endcsname\relax
  \def\citenamefont#1{#1}\fi
\expandafter\ifx\csname url\endcsname\relax
  \def\url#1{\texttt{#1}}\fi
\expandafter\ifx\csname urlprefix\endcsname\relax\def\urlprefix{URL }\fi
\providecommand{\bibinfo}[2]{#2}
\providecommand{\eprint}[2][]{\url{#2}}

\bibitem[{\citenamefont{Glashow}(1961)}]{Glashow:1961tr}
\bibinfo{author}{\bibfnamefont{S.}~\bibnamefont{Glashow}},
  \bibinfo{journal}{Nucl.Phys.} \textbf{\bibinfo{volume}{22}},
  \bibinfo{pages}{579} (\bibinfo{year}{1961}).

\bibitem[{\citenamefont{Salam}(1968)}]{sm_salam}
\bibinfo{author}{\bibfnamefont{A.}~\bibnamefont{Salam}},
  \emph{\bibinfo{title}{in Elementary Particle Theory, p. 367}}
  (\bibinfo{publisher}{Almquist and Wiksell}, \bibinfo{address}{Stockholm},
  \bibinfo{year}{1968}).

\bibitem[{\citenamefont{Weinberg}(1967)}]{PhysRevLett.19.1264}
\bibinfo{author}{\bibfnamefont{S.}~\bibnamefont{Weinberg}},
  \bibinfo{journal}{Phys. Rev. Lett.} \textbf{\bibinfo{volume}{19}},
  \bibinfo{pages}{1264} (\bibinfo{year}{1967}).

\bibitem[{\citenamefont{Higgs}(1964{\natexlab{a}})}]{Higgs:1964ia}
\bibinfo{author}{\bibfnamefont{P.~W.} \bibnamefont{Higgs}},
  \bibinfo{journal}{Phys. Lett.} \textbf{\bibinfo{volume}{12}},
  \bibinfo{pages}{132} (\bibinfo{year}{1964}{\natexlab{a}}).

\bibitem[{\citenamefont{Higgs}(1964{\natexlab{b}})}]{Higgs:1964pj}
\bibinfo{author}{\bibfnamefont{P.~W.} \bibnamefont{Higgs}},
  \bibinfo{journal}{Phys. Rev. Lett.} \textbf{\bibinfo{volume}{13}},
  \bibinfo{pages}{508} (\bibinfo{year}{1964}{\natexlab{b}}).

\bibitem[{\citenamefont{Guralnik et~al.}(1964)\citenamefont{Guralnik, Hagen,
  and Kibble}}]{Guralnik:1964eu}
\bibinfo{author}{\bibfnamefont{G.}~\bibnamefont{Guralnik}},
  \bibinfo{author}{\bibfnamefont{C.}~\bibnamefont{Hagen}}, \bibnamefont{and}
  \bibinfo{author}{\bibfnamefont{T.}~\bibnamefont{Kibble}},
  \bibinfo{journal}{Phys. Rev. Lett.} \textbf{\bibinfo{volume}{13}},
  \bibinfo{pages}{585} (\bibinfo{year}{1964}).

\bibitem[{\citenamefont{Englert and Brout}(1964)}]{Englert:1964et}
\bibinfo{author}{\bibfnamefont{F.}~\bibnamefont{Englert}} \bibnamefont{and}
  \bibinfo{author}{\bibfnamefont{R.}~\bibnamefont{Brout}},
  \bibinfo{journal}{Phys. Rev. Lett.} \textbf{\bibinfo{volume}{13}},
  \bibinfo{pages}{321} (\bibinfo{year}{1964}).

\bibitem[{\citenamefont{{The ALEPH, DELPHI, L3 and OPAL Collaborations, and the
  LEP Working Group for Higgs boson searches}}(2003)}]{Barate:2003sz}
\bibinfo{author}{\bibnamefont{{The ALEPH, DELPHI, L3 and OPAL Collaborations,
  and the LEP Working Group for Higgs boson searches}}},
  \bibinfo{journal}{Phys. Lett. B} \textbf{\bibinfo{volume}{565}},
  \bibinfo{pages}{61} (\bibinfo{year}{2003}).

\bibitem[{\citenamefont{{The CDF and D0 Collaborationsand the Tevatron New
  Physics and Higgs Working Group}}(2012)}]{TEVNPH:2012ab}
\bibinfo{author}{\bibnamefont{{The CDF and D0 Collaborationsand the Tevatron
  New Physics and Higgs Working Group}}} (\bibinfo{year}{2012}),
  \eprint{arXiv:1207.0449v2}.

\bibitem[{\citenamefont{Aad~{\it et al.}}(2012)}]{ATLAS:2012ae}
\bibinfo{author}{\bibfnamefont{G.}~\bibnamefont{Aad~{\it et al.}}}
  (\bibinfo{collaboration}{ATLAS Collaboration}) (\bibinfo{year}{2012}),
  \eprint{arXiv:1207.0319v1}.

\bibitem[{\citenamefont{Chatrchyan~{\it et
  al.}}(2012{\natexlab{a}})}]{Chatrchyan:2012tx}
\bibinfo{author}{\bibfnamefont{S.}~\bibnamefont{Chatrchyan~{\it et al.}}}
  (\bibinfo{collaboration}{CMS Collaboration}), \bibinfo{journal}{Phys. Lett.
  B} \textbf{\bibinfo{volume}{710}}, \bibinfo{pages}{26}
  (\bibinfo{year}{2012}{\natexlab{a}}), \eprint{arXiv:1202.1488}.

\bibitem[{\citenamefont{{G. Aad {\it et al.}}}(2012)}]{Atl:2012hzz}
\bibinfo{author}{\bibnamefont{{G. Aad {\it et al.}}}}
  (\bibinfo{collaboration}{Atlas Collaboration}), \bibinfo{journal}{Phys. Lett.
  B} \textbf{\bibinfo{volume}{710}}, \bibinfo{pages}{383}
  (\bibinfo{year}{2012}).

\bibitem[{\citenamefont{Chatrchyan~{\it et
  al.}}(2012{\natexlab{b}})}]{Chatrchyan:2012dg}
\bibinfo{author}{\bibfnamefont{S.}~\bibnamefont{Chatrchyan~{\it et al.}}}
  (\bibinfo{collaboration}{CMS Collaboration})
  (\bibinfo{year}{2012}{\natexlab{b}}), \eprint{1202.1997}.

\bibitem[{\citenamefont{Abulencia~{\it et al.}}(2007)}]{Abulencia:2005ix}
\bibinfo{author}{\bibfnamefont{A.}~\bibnamefont{Abulencia~{\it et al.}}}
  (\bibinfo{collaboration}{CDF Collaboration}), \bibinfo{journal}{J. Phys. G}
  \textbf{\bibinfo{volume}{34}}, \bibinfo{pages}{2457} (\bibinfo{year}{2007}).

\bibitem[{\citenamefont{{T. Aaltonen {\it et al.}}}(2012)}]{ZZXS}
\bibinfo{author}{\bibnamefont{{T. Aaltonen {\it et al.}}}}
  (\bibinfo{collaboration}{{CDF Collaboration}}), \bibinfo{journal}{{Phys. Rev.
  Lett.}} \textbf{\bibinfo{volume}{108}}, \bibinfo{pages}{101801}
  (\bibinfo{year}{2012}).

\bibitem[{\citenamefont{Sjostrand et~al.}(2006)\citenamefont{Sjostrand, Mrenna,
  and Skands}}]{Sjostrand:2006za}
\bibinfo{author}{\bibfnamefont{T.}~\bibnamefont{Sjostrand}},
  \bibinfo{author}{\bibfnamefont{S.}~\bibnamefont{Mrenna}}, \bibnamefont{and}
  \bibinfo{author}{\bibfnamefont{P.}~\bibnamefont{Skands}},
  \bibinfo{journal}{J. High Energy Phys.} \textbf{\bibinfo{volume}{05}},
  \bibinfo{pages}{026} (\bibinfo{year}{2006}).

\bibitem[{\citenamefont{Baur and Berger}(1993)}]{Baur:1992cd}
\bibinfo{author}{\bibfnamefont{U.}~\bibnamefont{Baur}} \bibnamefont{and}
  \bibinfo{author}{\bibfnamefont{E.~L.} \bibnamefont{Berger}},
  \bibinfo{journal}{Phys. Rev. D} \textbf{\bibinfo{volume}{47}},
  \bibinfo{pages}{4889} (\bibinfo{year}{1993}).

\bibitem[{\citenamefont{Lai~{\it et al.}}(2000)}]{Lai:1999wy}
\bibinfo{author}{\bibfnamefont{H.~L.} \bibnamefont{Lai~{\it et al.}}}
  (\bibinfo{collaboration}{CTEQ}), \bibinfo{journal}{Eur. Phys. J. C}
  \textbf{\bibinfo{volume}{12}}, \bibinfo{pages}{375} (\bibinfo{year}{2000}).

\bibitem[{\citenamefont{de~Florian and Grazzini}(2009)}]{deFlorian:2009hc}
\bibinfo{author}{\bibfnamefont{D.}~\bibnamefont{de~Florian}} \bibnamefont{and}
  \bibinfo{author}{\bibfnamefont{M.}~\bibnamefont{Grazzini}},
  \bibinfo{journal}{Phys. Lett. B} \textbf{\bibinfo{volume}{674}},
  \bibinfo{pages}{291} (\bibinfo{year}{2009}).

\bibitem[{\citenamefont{Anastasiou et~al.}(2009)\citenamefont{Anastasiou,
  Boughezal, and Petriello}}]{Anastasiou:2008tj}
\bibinfo{author}{\bibfnamefont{C.}~\bibnamefont{Anastasiou}},
  \bibinfo{author}{\bibfnamefont{R.}~\bibnamefont{Boughezal}},
  \bibnamefont{and}
  \bibinfo{author}{\bibfnamefont{F.}~\bibnamefont{Petriello}},
  \bibinfo{journal}{J. High Energy Phys.} \textbf{\bibinfo{volume}{04}},
  \bibinfo{pages}{003} (\bibinfo{year}{2009}).

\bibitem[{\citenamefont{Assamagan~{\it et al.}}(2004)}]{Assamagan:2004mu}
\bibinfo{author}{\bibfnamefont{K.}~\bibnamefont{Assamagan~{\it et al.}}}
  (\bibinfo{collaboration}{Higgs Working Group Collaboration}), pp.
  \bibinfo{pages}{1--169} (\bibinfo{year}{2004}), \eprint{hep-ph/0406152}.

\bibitem[{\citenamefont{Brein et~al.}(2004)\citenamefont{Brein, Djouadi, and
  Harlander}}]{Brein:2003wg}
\bibinfo{author}{\bibfnamefont{O.}~\bibnamefont{Brein}},
  \bibinfo{author}{\bibfnamefont{A.}~\bibnamefont{Djouadi}}, \bibnamefont{and}
  \bibinfo{author}{\bibfnamefont{R.}~\bibnamefont{Harlander}},
  \bibinfo{journal}{Phys. Lett. B} \textbf{\bibinfo{volume}{579}},
  \bibinfo{pages}{149} (\bibinfo{year}{2004}).

\bibitem[{\citenamefont{Ciccolini et~al.}(2003)\citenamefont{Ciccolini,
  Dittmaier, and Kramer}}]{Ciccolini:2003jy}
\bibinfo{author}{\bibfnamefont{M.~L.} \bibnamefont{Ciccolini}},
  \bibinfo{author}{\bibfnamefont{S.}~\bibnamefont{Dittmaier}},
  \bibnamefont{and} \bibinfo{author}{\bibfnamefont{M.}~\bibnamefont{Kramer}},
  \bibinfo{journal}{Phys. Rev. D} \textbf{\bibinfo{volume}{68}},
  \bibinfo{pages}{073003} (\bibinfo{year}{2003}).

\bibitem[{\citenamefont{Berger and Campbell}(2004)}]{Berger:2004pca}
\bibinfo{author}{\bibfnamefont{E.~L.} \bibnamefont{Berger}} \bibnamefont{and}
  \bibinfo{author}{\bibfnamefont{J.~M.} \bibnamefont{Campbell}},
  \bibinfo{journal}{Phys. Rev. D} \textbf{\bibinfo{volume}{70}},
  \bibinfo{pages}{073011} (\bibinfo{year}{2004}).

\bibitem[{\citenamefont{Campbell and Ellis}(2010)}]{SMXS}
\bibinfo{author}{\bibfnamefont{J.}~\bibnamefont{Campbell}} \bibnamefont{and}
  \bibinfo{author}{\bibfnamefont{R.~K.} \bibnamefont{Ellis}},
  \emph{\bibinfo{title}{{MCFM - Monte Carlo for FeMtobarn processes}}}
  (\bibinfo{year}{2010}), \bibinfo{note}{{We ran MCFM Monte Carlo with the
  MSTW2008 PDF set and varying the factorizzation and renormalizzation scale.}}

\bibitem[{\citenamefont{Baur et~al.}(1998)\citenamefont{Baur, Han, and
  Ohnemus}}]{Baur:1998}
\bibinfo{author}{\bibfnamefont{U.}~\bibnamefont{Baur}},
  \bibinfo{author}{\bibfnamefont{T.}~\bibnamefont{Han}}, \bibnamefont{and}
  \bibinfo{author}{\bibfnamefont{J.}~\bibnamefont{Ohnemus}},
  \bibinfo{journal}{Phys. Rev. D} \textbf{\bibinfo{volume}{57}},
  \bibinfo{pages}{2823} (\bibinfo{year}{1998}).

\bibitem[{\citenamefont{Brun et~al.}()\citenamefont{Brun, Hagelberg, Hansroul,
  and Lassalle}}]{GEANT}
\bibinfo{author}{\bibfnamefont{R.}~\bibnamefont{Brun}},
  \bibinfo{author}{\bibfnamefont{R.}~\bibnamefont{Hagelberg}},
  \bibinfo{author}{\bibfnamefont{M.}~\bibnamefont{Hansroul}}, \bibnamefont{and}
  \bibinfo{author}{\bibfnamefont{J.}~\bibnamefont{Lassalle}},
  \eprint{CERN-DD-78-2-REV and CERN-DD-78-2}.

\bibitem[{\citenamefont{Acosta~{\it et al.}}(2002)}]{Acosta:2002hx}
\bibinfo{author}{\bibfnamefont{D.}~\bibnamefont{Acosta~{\it et al.}}},
  \bibinfo{journal}{Nucl. Instrum. Meth. A} \textbf{\bibinfo{volume}{494}},
  \bibinfo{pages}{57} (\bibinfo{year}{2002}).

\bibitem[{\citenamefont{Catani and Grazzini}(2007)}]{Catani:2007vq}
\bibinfo{author}{\bibfnamefont{S.}~\bibnamefont{Catani}} \bibnamefont{and}
  \bibinfo{author}{\bibfnamefont{M.}~\bibnamefont{Grazzini}},
  \bibinfo{journal}{Phys. Rev. Lett.} \textbf{\bibinfo{volume}{98}},
  \bibinfo{pages}{222002} (\bibinfo{year}{2007}).

\bibitem[{\citenamefont{Grazzini}(2008)}]{Grazzini:2008tf}
\bibinfo{author}{\bibfnamefont{M.}~\bibnamefont{Grazzini}},
  \bibinfo{journal}{J. High Energy Phys.} \textbf{\bibinfo{volume}{0802}},
  \bibinfo{pages}{043} (\bibinfo{year}{2008}).

\bibitem[{\citenamefont{TeV4LHC}()}]{TeV4LHC}
\bibinfo{author}{\bibnamefont{TeV4LHC}}, \emph{\bibinfo{title}{Standard model
  higgs cross sections at hadron colliders}},
  \bibinfo{note}{{http://maltoni.home.cern.ch/maltoni/TeV4LHC/SM.html}}.

\bibitem[{\citenamefont{Campbell et~al.}()\citenamefont{Campbell, Ellis, and
  Williams}}]{mcfm}
\bibinfo{author}{\bibfnamefont{J.}~\bibnamefont{Campbell}},
  \bibinfo{author}{\bibfnamefont{K.}~\bibnamefont{Ellis}}, \bibnamefont{and}
  \bibinfo{author}{\bibfnamefont{C.}~\bibnamefont{Williams}},
  \emph{\bibinfo{title}{Monte carlo for femtobarn processes}},
  \bibinfo{note}{{http://mcfm.fnal.gov/mcfm.pdf}}.

\bibitem[{\citenamefont{Campbell and Ellis}(1999)}]{Campbell:1999ah}
\bibinfo{author}{\bibfnamefont{J.~M.} \bibnamefont{Campbell}} \bibnamefont{and}
  \bibinfo{author}{\bibfnamefont{R.~K.} \bibnamefont{Ellis}},
  \bibinfo{journal}{Phys. Rev. D} \textbf{\bibinfo{volume}{60}},
  \bibinfo{pages}{113006} (\bibinfo{year}{1999}).

\bibitem[{\citenamefont{Beringer~{\it et al.}}(2012)}]{Beringer:2012}
\bibinfo{author}{\bibfnamefont{J.}~\bibnamefont{Beringer~{\it et al.}}}
  (\bibinfo{collaboration}{Particle Data Group}), \bibinfo{journal}{Phys. Rev.
  D} \textbf{\bibinfo{volume}{86}}, \bibinfo{pages}{010001}
  (\bibinfo{year}{2012}).

\bibitem[{\citenamefont{Aaltonen et~al.}(2011)}]{Aaltonen:2011zq}
\bibinfo{author}{\bibfnamefont{T.}~\bibnamefont{Aaltonen}} \bibnamefont{et~al.}
  (\bibinfo{collaboration}{CDF}), \bibinfo{journal}{Phys. Rev. D}
  \textbf{\bibinfo{volume}{83}}, \bibinfo{pages}{112008}
  (\bibinfo{year}{2011}), \eprint{1102.4566}.

\end{thebibliography}

\end{document}